\documentclass{mufson}

\def\etal {{\it et al.}}

\begin{document}

\title{The Search for Neutrino-Antineutrino Mixing from Lorentz Invariance Violation using Neutrino Interactions in MINOS}

\author{S.~ Mufson$^*$}

\address{Astronomy Department, Indiana University, USA\\
Bloomington, IN 47405\\
$^*$E-mail: mufson@astro.indiana.edu}

\author{B.~Rebel}

\address{Fermi National Accelerator Laboratory,\\
Batavia, Illinois 60510, USA\\
E-mail: brebel@fnal.gov}

\begin{abstract}
We searched for a sidereal modulation in the rate of neutrinos observed by the MINOS far detector.  The detection of these signals could be a signature of neutrino-antineutrino mixing due to Lorentz and CPT violation as described by the Standard-Model Extension framework.  We found no evidence for these sidereal signals and we placed limits on the coefficients in this theory describing the effect.  
\end{abstract}

\bodymatter

\section{Introduction}
In the SME, LV and CPTV could manifest themselves at observable energies through a dependence of the neutrino oscillation probability on the direction of neutrino propagation with respect to the Sun-centered inertial frame.   An experiment that has both its neutrino beam and detector fixed on the Earth's surface could then observe a sidereal modulation in the number of neutrinos detected from the beam.  MINOS is such an experiment.

Denote the usual neutrino survival probability in the two-flavor approximation as $P^{(0)}_{\nu_\mu \rightarrow \nu_\mu} \approx 1-\sin^2{(2 \theta_{23})} \sin^2{(1.27 \Delta m^2_{32} L /E)}$, where $\theta_{23}$ is the angle describing mixing between the second and third mass states and $\Delta m^{2}_{32}$ is the difference in the squares of those mass states.  The energy of the neutrino is $E$ and the distance it travels is $L$.  Then LV and CPTV  that would cause lepton-number violating mixing between neutrinos and antineutrinos introduces an additional perturbation term to the survival probability~\cite{dkm},
\begin{equation}
\label{eq:totalprob}
P_{\nu_\mu \rightarrow \nu_\mu} = P^{(0)}_{\nu_\mu \rightarrow \nu_\mu} + P^{(2)}_{\nu_\mu \rightarrow \nu_\mu},
\end{equation}
where the perturbation term $P^{(2)}_{\nu_\mu \rightarrow \nu_\mu}$ can be written~\cite{dkm}
\begin{eqnarray}
\label{eq:osc}
P^{(2)}_{\nu_\mu \rightarrow \nu_\mu} &=& L^2 \bigl\{P_{\mathcal{C}} \nonumber \\ &+&
 P_{\mathcal{A}_{s}}\sin\omega_{\oplus}T_{\oplus} + P_{\mathcal{A}_{c}}\cos\omega_{\oplus}T_{\oplus} \\ \nonumber
 &+& P_{\mathcal{B}_{s}}\sin2\omega_{\oplus}T_{\oplus} + P_{\mathcal{B}_{c}}\cos2\omega_{\oplus}T_{\oplus} \\ \nonumber
 &+& P_{\mathcal{D}_{s}}\sin3\omega_{\oplus}T_{\oplus} + P_{\mathcal{D}_{c}}\cos3\omega_{\oplus}T_{\oplus} \\ \nonumber
&+& P_{\mathcal{F}_{s}}\sin4\omega_{\oplus}T_{\oplus} + P_{\mathcal{F}_{c}}\cos4\omega_{\oplus}T_{\oplus} 
\bigr\}.\nonumber 
\end{eqnarray}
Here $L = 735$~km is the distance from neutrino production in the NuMI beam to the MINOS FD~\cite{minoscc}, $\omega_\oplus= 2 \pi/(23^h 56^m 04.0982^s)$ is the Earth's sidereal frequency, and $T_\oplus$ is the local sidereal arrival time of the neutrino event.  Eq.(\ref{eq:osc}) shows that harmonic variations at the sidereal frequency are visible up to $4 \omega_\oplus$.  The $P$ parameters contain the LV and CPTV information on neutrino-antineutrino mixing.  They depend on the SME coefficients $\tilde H^{\,\alpha}_{a \bar b}$ and $\tilde g^{\,\alpha\beta}_{a \bar b}$, the neutrino energy, and the direction of the neutrino propagation in a coordinate system fixed on the rotating Earth~\cite{dkm}. 

A more complete description of this work can be found in~\cite{rebel}.

\section{Data Analysis}

This analysis uses a data set of neutrino interactions acquired by MINOS from May, 2005 through April, 2012.  
The interactions were selected using standard MINOS criteria for beam and data quality~\cite{cuts}.  In addition, the events were required to interact within the 4.2~kiloton FD fiducial volume.  This selection enables MINOS to establish each event as a CC $\nu_\mu$ interaction by identifying the outgoing $\mu^-$.  We focused on CC events to maximize the $\nu_\mu$ disappearance signal.  There are a total of 2,463 CC events in this analysis.   

\begin{figure}[h]
\centerline{\includegraphics[width=3.25in]{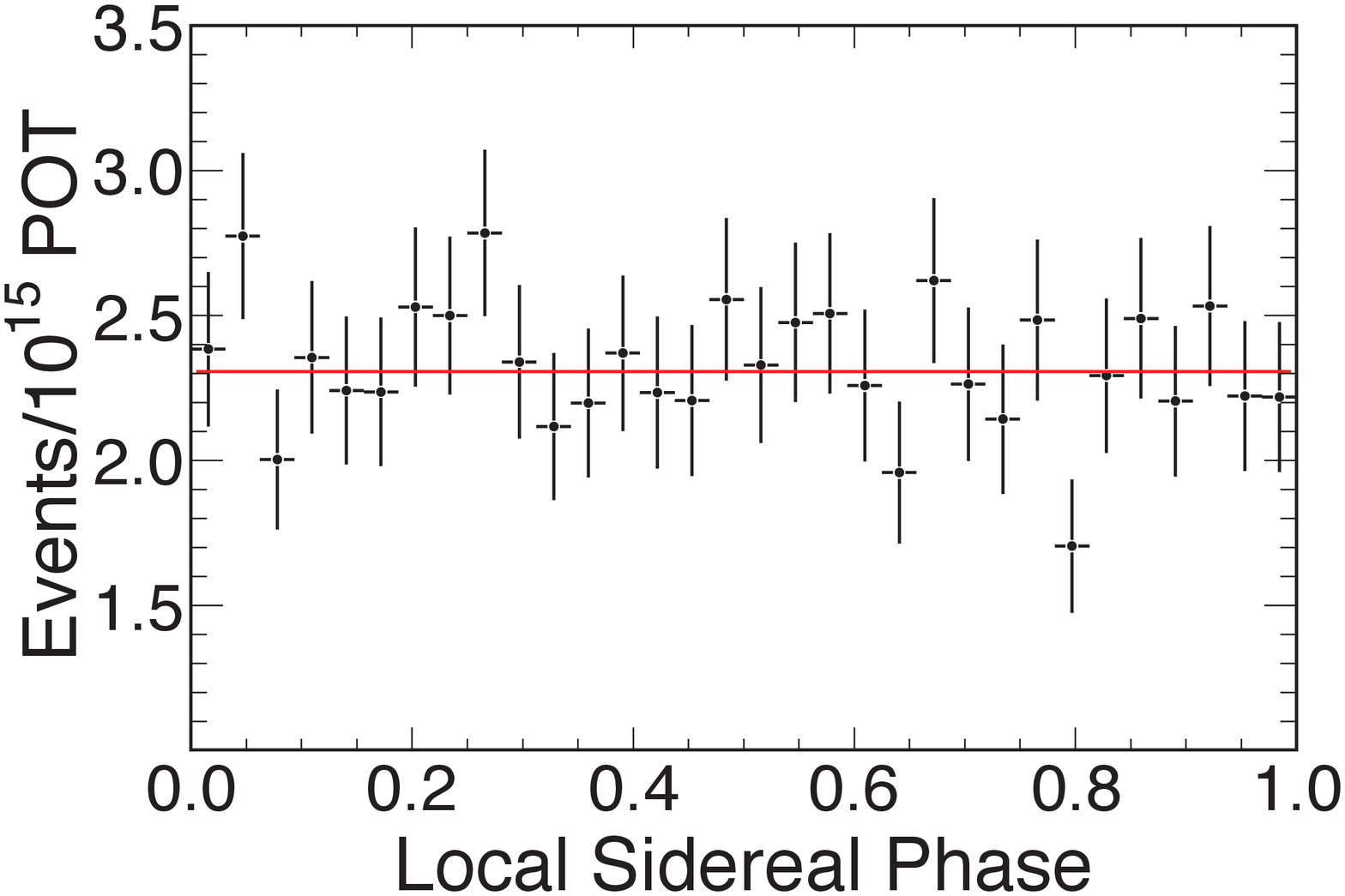}}
\caption{\label{fig:data_rate} The sidereal phase diagram for the CC neutrino rate for the FD data.  The mean rate of $2.31 \pm 0.05$ events per $10^{15}$~POT is superposed.  $\chi^{2}/ndf = 23.7/31$ for the fit.}
\end{figure}

Figure~\ref{fig:data_rate} shows the rate histogram in sidereal phase for the events in our analysis.
The statistically significant fit to a constant rate implies there are no sidereal modulations in the data sample.  

We performed an FFT on the rate histogram in Fig.~\ref{fig:data_rate} and computed the power in the harmonic terms $\omega_{\oplus}T_{\oplus}, \ldots, 4\omega_{\oplus}T_{\oplus}$ appearing in the oscillation probability, Eq.(\ref{eq:osc}).   
The results of the FFT analysis are given in Table~\ref{table:dataPower}.  
\begin{table}[h]
\tbl{Results of the FFT analysis.   $\cal{P}_F$ is the probability that the power is a noise fluctuation. }
{\begin{tabular}{@{}ccc@{}}\toprule
 & Harmonic Power(FFT)& ~~~~~$\cal{P}_F$~~~~~ \\ \colrule
$\omega_{\oplus}T_{\oplus}$ &  0.928 & 0.65\\ 
$2\omega_{\oplus}T_{\oplus}$  & 0.574 & 0.89 \\
$3\omega_{\oplus}T_{\oplus}$  & 1.388 & 0.48 \\
$4\omega_{\oplus}T_{\oplus}$  & 1.223 & 0.53 \\\botrule
\end{tabular} 
}
\label{table:dataPower}\end{table}
In Rebel and Mufson~\cite{rebel} we determine the statistical significance of the harmonic powers in Table~\ref{table:dataPower} by simulation.  We find that 3.45 is the 99.7\% confidence level (C.L.) for the probability that a measured quadratic sum of powers for any harmonic was {\it not} drawn from a distribution having a sidereal signal. 
Since none of the harmonic powers  exceed our  99.7\%(FFT) detection threshold, we conclude that there is no evidence for a sidereal modulation resulting from mixing between neutrinos and antineutrinos as predicted by Eq.~(\ref{eq:osc}) in this neutrino data set.  

We investigated the sensitivity of our results to several sources of systematic uncertainties.  We found that systematics were unimportant.

\section{Limits}

We determined the confidence limits for the SME coefficients by methods we developed in our previous papers~\cite{minoscc, paper2, paper3}.  In these limit calculations the odd harmonics in  eq.(\ref{eq:osc}) vanish.  Since we set all but one coefficient to zero in computing the limits and the odd harmonics involve products of different SME coefficients, they disappear from  eq.(\ref{eq:osc}).
We start by simulating a set of experiments in which there is no sidereal modulation.  Each of the simulated experiments contain the same number of interactions as the data set.  We then introduce an infinitesimal LV sidereal signal into the experiments with the one nonzero SME coefficient.  We then generate the sidereal phase diagrams for these experiments and test whether there is an LV signal detectable in the phase diagrams with the same FFT analysis used to analyze the data in Fig.~\ref{fig:data_rate}. 

In our simulation, neutrinos are generated by modeling the NuMI beam line, including hadron production by the 120 GeV$/c$ protons striking the target and the propagation of the hadrons through the focusing elements and decay pipe to the beam absorber.  The simulation then calculates the probability that any neutrinos generated traverse the FD.  The FD neutrino event simulation takes the neutrinos from the NuMI simulation, along with weights determined by decay kinematics, and uses this information as input into the simulation of the interactions in the FD.   
We inject a sidereal signal in the simulation by calculating the survival probability for each simulated neutrino based on the even harmonics in Eq.~(\ref{eq:osc}) using a chosen value for the magnitude of the nonzero SME coefficient, the energy of the simulated neutrino, and the distance the neutrino travels to the FD. 

The 99.7\% C.L. limits on the 66 $\tilde H^{\,\alpha}_{a \bar b}$ and $\tilde g^{\,\alpha\beta}_{a \bar b}$ SME coefficients are found in tables in Rebel and Mufson~\cite{rebel}.    
For these tables, we repeated the simulation for each coefficient 250 times and averaged the results.

\section*{Acknowledgments}

This work was supported in part by the Indiana University Center for Spacetime Symmetries (IUCSS) and by the U.S. Department of Energy Office of Science with grant DE-FG02-91ER40661.


\begin{thebibliography}{xx}

\bibitem{dkm}
J.S.\ D\'iaz, V.A.\  Kosteleck\'y, and M.\ Mewes, 
Phys.\ Rev.\ D\ {\bf 80}, 076007 (2009).

\bibitem{minoscc}
MINOS Collaboration,
P.\ Adamson \etal,
Phys.\ Rev.\ Lett.\ {\bf 101}, 151601 (2008).

\bibitem{cuts}
MINOS Collaboration,
P.\ Adamson \etal,
Phys.\ Rev.\ Lett.\ {\bf 106}, 181801 (2011).

\bibitem{rebel}
B.\ Rebel and S.\ Mufson
AstroPart.Phys, in press (2013) arXiv:1301.4684.

\bibitem{paper2}
MINOS Collaboration,
P.\ Adamson \etal,
Phys.\ Rev.\ Lett.\ {\bf 105}, 151601 (2010).

\bibitem{paper3}
MINOS Collaboration,
P.\ Adamson \etal,
Phys.\ Rev.\ D\ {\bf 85}, 031101 (2012).


\end{thebibliography}
\end{document}